\documentclass[12pt,a4paper]{article}

\usepackage{a4wide}
\usepackage{latexsym}
\usepackage{epsf}
\usepackage{amssymb}
\usepackage{graphicx}
\usepackage{amsmath, cite}
\usepackage{amsmath,amssymb,amsthm}
\usepackage{verbatim}
\usepackage{hyperref}

\renewcommand{\d}{\textrm{d}}
\newcommand{\e}{\textrm{e}}

\newcommand{\SL}{\mathop{\rm SL}}
\newcommand{\SO}{\mathop{\rm SO}}
\newcommand{\SU}{\mathop{\rm SU}}

\newcommand{\be}{\begin{equation}}
\newcommand{\ee}{\end{equation}}
\newcommand{\ba}{\begin{eqnarray}}
\newcommand{\ea}{\end{eqnarray}}

\newcommand{\lp}{\left(}
\newcommand{\rp}{\right)}

\renewcommand{\d}{\textrm{d}}

\begin{document}
\numberwithin{equation}{section}

\begin{center}

\begin{flushright}
{\small UUITP-26/12
 \\  MAD-TH-12-10\\
 SU/ITP-12/44} \normalsize
\end{flushright}
\vspace{0.3 cm}

{\LARGE \bf{A note on obstinate tachyons\\\vspace{0.2cm} in classical dS solutions}}

\vspace{1.1 cm} {\large  U.H. Danielsson$^a$, G. Shiu$^b$, T. Van Riet$^c$, T. Wrase$^d$}\\

\vspace{0.8 cm}{$^{a}$ Institutionen f{\"o}r fysik och astronomi,\\ Uppsala Universitet, Uppsala, Sweden}\\
\vspace{.1 cm} {$^b$ Department of Physics, University of Wisconsin, \\Madison, WI 53706, USA
%\emph{and}
  \\
  Department of Physics \& Institute for Advanced Study, \\Hong Kong University of Science and Technology, Hong Kong}\\
\vspace{.1 cm} {$^c$ Instituut voor Theoretische Fysica, K.U. Leuven,\\ Celestijnenlaan 200D B-3001 Leuven, Belgium}\\
\vspace{.1 cm} {$^d$ Stanford Institute for Theoretical Physics,\\ Stanford University, Stanford, CA 94305, USA
\footnote{{\ttfamily ulf.danielsson @ physics.uu.se, shiu @ physics.wisc.edu, thomasvr @ itf.fys.kuleuven.be,  timm.wrase @ stanford.edu}
} }

\vspace{0.8cm}

{\bf Abstract}
\end{center}

\begin{quotation}
The stabilisation of the dilaton and volume in tree-level flux compactifications leads to model independent and thus very powerful existence and stability criteria for dS solutions. In this paper we show that the sizes of cycles wrapped by orientifold planes are scalars whose scalings in the potential are not entirely model independent, but enough to entail strong stability constraints. For all known dS solutions arising from massive IIA supergravity flux compactifications on $\SU(3)$-structure manifolds the tachyons are exactly within the subspace spanned by the dilaton, the total volume and the volumes of the orientifold cycles. We illustrate this in detail for the well-studied case of the $O6$ plane compactification on $\SU(2)\times\SU(2)/\mathbb{Z}_2\times \mathbb{Z}_2$. For that example we uncover another novel structure in the tachyon spectrum: the dS solutions have a singular, but supersymmetric, Minkowski limit, in which the tachyon exactly aligns with the sgoldstino.

\end{quotation}

\newpage

\section{Introduction}
Understanding the structure of non-supersymmetric vacua in low-energy string theory is a most relevant but challenging question, especially when it concerns de Sitter vacua. The current situation can roughly be summarised as follows: All constructions of meta-stable de Sitter vacua in string theory (most prominently \cite{Kachru:2003aw, Saltman:2004sn, Balasubramanian:2005zx, Lebedev:2006qq, Rummel:2011cd}) are fairly involved since they always make use of non-perturbative corrections. Sometimes perturbative corrections are necessary as well and the construction proceeds in two steps: First a stable AdS vacuum is found that is then uplifted to a dS solution. The reason for this is that the authors of \cite{Kachru:2003aw,Saltman:2004sn, Balasubramanian:2005zx, Lebedev:2006qq, Rummel:2011cd} (and many others) aim at finding semi-realistic de Sitter solutions, which for instance should have a built-in mechanism that allows for a tunably small size of the cosmological constant. However, the involvedness of the construction has lead to various concerns in the literature regarding their consistency, see for instance \cite{Banks:2004xh, Komargodski:2009pc, Blaback:2012nf, Conlon:2012tz, Bena:2012bk, Banks:2012hx }. Therefore it would be desirable to obtain a simple, fully explicit dS vacuum. If such vacua can be found at the 10D supergravity tree-level, then we should have full control over the constructions.

It is well understood that such constructions necessarily involve singular sources in order to evade no-go theorems \cite{deWit:1986xg}. Many attempts to construct such vacua at tree-level have appeared over the last five years \cite{ Flauger:2008ad, Caviezel:2008tf, Danielsson:2009ff, deCarlos:2009fq,deCarlos:2009qm, Caviezel:2009tu , Danielsson:2010bc, Andriot:2010ju, Dong:2010pm,  Dibitetto:2011gm , Shiu:2011zt, Danielsson:2011au, Danielsson:2012by}, inspired by the original papers \cite{Hertzberg:2007wc, Silverstein:2007ac}. If one restricts oneself to geometric fluxes and non-exotic branes, in order to trust the supergravity limit, one can show that the solutions necessarily involve negative tension orientifold planes, with a minimal set of non-zero fluxes \cite{ Haque:2008jz,  Danielsson:2009ff, Wrase:2010ew, VanRiet:2011yc}. Despite all the attempts the models suffer from the following shortcomings, that need to be dealt with in order to progress:
\medskip

1. All the geometric orientifold compactifications that allow for an explicit computation of all moduli masses have tachyons \cite{Flauger:2008ad, Caviezel:2008tf,  deCarlos:2009fq, deCarlos:2009qm, Caviezel:2009tu, Danielsson:2010bc, Dibitetto:2011gm, Danielsson:2011au}. This problem is the focus of this paper. If, however, non-geometric fluxes are allowed meta-stable dS solutions  can be found \cite{deCarlos:2009fq, deCarlos:2009qm, Danielsson:2012by, Blaback:2013ht, Damian:2013dq}. It is unclear whether these solutions can really be trusted as proper string theory solutions.
\medskip

2.  The orientifold sources are smeared instead of localised, because the smeared limit allows the construction of solutions at the level of the 10-dimensional supergravity equations of motion and, only when the orientifolds are smeared do we obtain gauged supergravities after compactification (see for instance \cite{Roest:2009dq}). It was first discussed in \cite{Douglas:2010rt} that the smeared limit might lead to ``fake'' solutions in the sense that there is no solution anymore once the sources are localised. This was elaborated upon in \cite{Blaback:2010sj, Blaback:2011nz, Blaback:2011pn} where it was demonstrated that the smearing procedure seems harmless for BPS solutions, but is indeed subtle for non-BPS solutions. In particular, it was shown that the localised solutions have certain singular fluxes  that might invalidate the solution. In the BPS cases the consistency of the smearing limit has only been verified for solutions with sources that are all parallel \cite{Blaback:2010sj}, whereas the solutions studied in this paper have intersecting O6 planes, for which no conclusive arguments exist, even in the BPS case. Reference \cite{Saracco:2012wc} was able to derive part of the would-be solution with such intersecting O6 planes, whereas reference \cite{ McOrist:2012yc} provides some arguments against their existence. It would be most interesting to settle this issue.

\medskip
3. It is not entirely clear how one can understand these solutions from the point of view of perturbative string theory. For example when the Romans mass in IIA supergravity is non-zero we do not have a perturbative string theory description and hence it is unclear how one should define $O6$ planes. In  general, if singular sources are present, it is desirable to have a smooth M/F-theory lift, which can put the solution on a firm basis. This is especially problematic for $O6$ planes in massive IIA supergravity \cite{Saracco:2012wc, McOrist:2012yc}. One reason to be suspicious about these issues is  the fact that $O6$ planes with Romans mass give flux solutions which are radically different from any other orientifold flux compactification (such as those of\cite{Dasgupta:1999ss}): they are claimed to allow full moduli stabilisation at the classical level, with unbounded $F_4$ fluxes and a tunable hierarchy between the AdS curvature and the KK scale \cite{Derendinger:2004jn, Villadoro:2005cu, DeWolfe:2005uu, Ihl:2006pp, Ihl:2007ah, Caviezel:2008ik}\footnote{Interestingly the same kind of solutions can be found in the `formally T-dual' setting of O5/O7 compactifications in IIB \cite{Caviezel:2009tu}, where some of these difficulties might be absent.}.  Although, as stressed in \cite{ McOrist:2012yc}, it is unclear whether these properties remain when we the sources are fully  localised.
At the same time these are exactly the attractive properties, which we try to extend to dS vacua as well.

\medskip
In this paper we put these last two issues aside and take a pragmatic viewpoint. The massive IIA $O6$ compactifications give rise to consistent four-dimensional supergravities and we take a closer look at the obstinate tachyons that were present in all geometric dS compactifications. A large scan for such solutions was performed in \cite{Danielsson:2011au}. For all numerical solutions\footnote{Recently some powerful algebraic geometry methods haven been invented to minimize the scalar potentials and to improve on the search for vacua \cite{Mehta:2012wk}.} the spectrum contained at least one tachyon. The effective theories, in which the tachyons were computed, are $\mathcal{N}=1$ supergravities.  These supergravities, coming from orientifolded twisted tori, range from having 6 to 14 real scalar fields. The more moduli, the more difficult it becomes to have meta-stable solutions and this could be a possible explanation for the negative results so far \cite{Marsh:2011aa, Chen:2011ac, Sumitomo:2012wa, Bachlechner:2012at, Sumitomo:2012vx, Sumitomo:2012cf }. For example in \cite{Marsh:2011aa} it was argued that in random $\mathcal{N}=1$ dS extrema arising from an F-term potential 5\%-15\% of the moduli are tachyonic. One also expects on general grounds \cite{Aazami:2005jf, Chen:2011ac} that the likelihood for a dS critical point to be actually a dS minimum goes like $e^{-c N^2}$, where $N$ is the number of scalar fields and $c$ a model dependent constant. Both of these facts seem pretty much consistent with the explicit constructions of dS extrema in type IIA that all have one or a few tachyons. However, the constant $c$ for random $\mathcal{N}=1$ supergravities is substantially smaller than unity so that dS minima for cases with 6 to 14 real moduli are not that unlikely, in particular if one neglects flux quantization which then leads to infinite families of dS extrema. As is explicitly shown in \cite{Danielsson:2012by} the resulting number of meta-stable dS in a simple model with generalized fluxes is substantially smaller than expect from random matrix theory arguments. It is therefore likely that there is a structural reason for these tachyons.  We are aware of two structural reasons for tachyons in the literature:

\medskip
1. Similar to no-go theorems for the existence of dS critical points, there exist no-go theorems for dS critical points that are meta-stable \cite{Shiu:2011zt,VanRiet:2011yc}. These no-go theorems are based on the scaling of the potential with the string coupling and the volume. These scalings are model independent and just depend on the rank of the flux and the kind of sources that are involved in the compactification. Therefore one can show the existence of tachyons for entire classes of models since the details of the models are not relevant for this. In this paper this idea will be generalised.

\medskip
2. In compactifications that lead to four-dimensional supergravity theories one can identify the field directions, which are most likely to become unstable when SUSY is broken by a small amount (where small means compared to the scale set by the massive scalars in the SUSY vacuum). This direction is the sgoldstino direction and there are circumstance in which one can demonstrate that it is necessarily tachyonic \cite{GomezReino:2006dk, Covi:2008ea, GomezReino:2008bi, Borghese:2010ei}.

\medskip
As we demonstrate in this paper there is evidence that all tachyons found so far in tree-level dS solutions are due to the first quoted scaling arguments. In the single case we know of where SUSY can be slightly broken in these dS compactifications, the sgoldstino is indeed the tachyon. This occurs for the prime example of the $\SU(2)\times\SU(2)$ compactification of massive IIA supergravity with four intersecting $O6$ planes. Therefore these two structural reasons are connected in this example and below we discuss in more detail how this happens.

The rest of this paper is organised as follows. In section \ref{semi} we generalise the existing stability constraints from scaling arguments by demonstrating that the scaling of the potential with respect to the orientifold volume is also very constraining. In section \ref{SU2} we illustrate this  in detail for the case of $\SU(2)\times \SU(2)$. Subsequently we demonstrate how the tachyon in the `semi'-universal scaling directions  (the dilaton, the 6D volume and the orientifold volumes)  connects nicely to the sgoldstino tachyon when SUSY is slightly broken. In section \ref{scan} we then present the numerical results for the scan of dS solutions in massive IIA performed in \cite{Danielsson:2011au}. The numerics demonstrate that all of the solutions studied in the scan have tachyons in the directions that scale `semi'-universal.  Section \ref{discussion} is a brief discussion of results and future directions. Finally we end with a technical  appendix that demonstrates that the solution on $\SU(2)\times \SU(2)$ that possesses some isotropic symmetry in field space, cannot be extended to more general $\SU(3)$-structures, thereby resolving an issue of \cite{Danielsson:2010bc}.

\section{A ``semi-universal'' modulus} \label{semi}

\subsection{The universal moduli: volume and string coupling}
Any compactification of 10-dimensional supergravity down to $D$ spacetime dimensions involves at least two moduli, the string coupling, $g_s = \e^{\phi}$, and the volume modulus, $\rho$, of the internal space. In what follows we will use the notation of \cite{Hertzberg:2007wc} and write the 10-dimensional metric in string frame as follows
\begin{equation}
\d s^2_{10} = \tau^{-2}\d s_D^2 + \rho\,\,\d s^2_{10-D}\,,
\end{equation}
where $\tau$ and $\rho$ are the two universal moduli: $\rho$ measures the internal volume (in string frame) and $\tau$ is the following combination of string coupling and volume:
\begin{equation}
\tau^{D-2} = \e^{-2\phi}\,\rho^{\frac{10-D}{2}}\,.
\end{equation}
This combination is such that the $D$-dimensional metric is in $D$-dimensional Einstein frame, as it should be. In the above metric we have ignored the warpfactor. This corresponds to the approximation in which the brane and orientifold sources are smeared. We refer to \cite{Douglas:2010rt, Blaback:2010sj, Blaback:2011nz} for possible caveats of this approximation.

At tree-level the potential energy consists of various terms that originate from the RR fluxes $F_i$, denoted $V_i$, the $H$ flux, denoted $V_H$, the internal curvature, denoted $V_R$, and the $Dp/Op$ tension, denoted $V_{Op/Dp}$. Each of these ingredients scales with the \emph{universal} moduli in a \emph{universal} way as follows:
\begin{align}
& V_R \sim - R_{\text{internal}}\rho^{-1}\tau^{-2}\,,\\
& V_H \sim |H|^2 \rho^{-3}\tau^{-2}\,,\\
& V_{Op/Dp} \sim T_p\,\rho^{\frac{2p-D-8}{4}}\,\tau^{-\frac{D+2}{2}}\,,\\
& V_{i} \sim |F_i|^2\, \rho^{\frac{10-D}{2}-i}\,\tau^{-D}\,,
\end{align}
where $T_p$ is the net $Dp+Op$ tension. It has been shown in \cite{Shiu:2011zt} (for $D=4$)  and \cite{VanRiet:2011yc} (for $D>4$) that dS solutions that are meta-stable in the $\rho, \tau$ plane necessarily have net negative orientifold tension and negative internal curvature\footnote{In reference \cite{Shiu:2011zt} the $O3$ and $O4$ models seem to evade this, but these models then necessarily involve fluxes that are usually projected out by the orientifold planes. There could exist more exotic orientifolds that are defined through involutions that act only on localised cycles, such as the twisted sector in orbifold models, for which the bulk fluxes are then not projected out.}
\begin{equation}
\text{dS solutions}\qquad \rightarrow \qquad V_R>0\,,\quad V_{Op/Dp}<0\,.
\end{equation}
For this reason we will shorten the notation for the source energy to $V_{Op}$.

\subsection{The orientifold volume}
The required presence of an orientifold plane is what allows us to define a third modulus, with almost universal scalings as we will now describe. This modulus, which we denote $\sigma$, is the volume modulus of the submanifold wrapped by the orientifold plane. In case there are multiple, intersecting sources, there can be multiple $\sigma$-moduli, denoted $\sigma_a$.
We write the internal metric as
\begin{equation}\label{sigma}
\d s^2_{10-D} = \rho\,\, \Bigl(\,\,\sigma^A\underbrace{\d
s^2_{p+1-D}}_{\text{$\parallel$ $Op$}}\quad + \quad \sigma^B\underbrace{\d
s^2_{9-p}}_{\text{$\perp$ $Op$}}\,\,\Bigr)\,,
\end{equation}
where $A, B$ are constants that are determined below. The first part in the metric is along the $Op$ plane and the second is
transverse. For simplicity we did not consider cross terms. To make $\sigma$ independent of the $\rho$-scaling one has to  make sure that $\sigma$ drops out of the determinant of the internal metric.  Hence
\begin{equation}
A (p+1-D) + B (9-p) =0 \,.
\end{equation}
Since we will not be bothered with any normalisation of the scalars
we solve this equation with the simple choice
\begin{equation}\label{eq:AB}
A = p-9\,,\qquad B = p+1-D\,.
\end{equation}
The dependence of $V_{Op}$ on $\sigma$ can readily be
computed
\begin{equation}
 V_{Op} \sim T_p\,
\rho^{\frac{2p-D-8}{4}}\,\tau^{-\frac{D+2}{2}}\,\sigma^{\tfrac{1}{2}(p-9)(p+1-D)}\,.
\end{equation}

The dependence of the flux energies $V_i$ and $V_H$ is also given by simple scalings, but the expressions are semi-universal (semi-model independent) in the sense that one has to specify how many legs the flux has alongside the $Op$ plane. Interestingly, the parity rules for the fluxes (see \cite{VanRiet:2011yc} for a simple summary), restrict the possible legs strongly. As an example, let us consider $O6$ planes in IIA for $D=4$ compactifications. Then we have that the Romans mass $F_0$ is even, $F_2$ is odd, $F_4$ is even, $F_6$ is odd. So $F_2$, $F_4$ and $F_6$ all have half their legs along the $O6$ plane and half their legs transverse to it. Remarkably this implies that none of the RR forms have an energy dependence on $\sigma$.  This is due to the fact that $A = - B$ when $p=6$, $D=4$. Hence we have
\begin{equation}\label{eq:RRscaling}
V_{i} \sim |F_i|^2\, \rho^{3-i}\,\tau^{-4}\,\,,
\end{equation}
Since $H$ is odd it can have either one or three legs transverse to the $O6$ plane and therefore has two dependencies:
\begin{equation}\label{eq:Hscaling}
V_H \sim  \rho^{-3}\tau^{-2}\,\Bigl( |H_1|^2 \sigma^{3} +  |H_3|^2 \sigma^{-9}\Bigr)\,.\\
\end{equation}
The dependence of the curvature energy $V_R$ on $\sigma$ is the most model dependent part and is more involved and is discussed in the next section.

There is a notable subtlety to this approach. It is possible to construct orientifold compactifications in which the volume of the orientifold cycle is determined by the overall volume $\rho$. In other words, there exist situations in which $\sigma$ is not a dynamical field. Such examples appeared in the very first attempts to construct de Sitter solutions at the classical level \cite{Silverstein:2007ac, Haque:2008jz}. In these examples the internal space $M_6$ is a direct product of two 3-manifolds $M_3$ and the orientifold involution exchanges the two $M_3$'s:
\begin{equation}\label{dynamical}
M_6 = \frac{M_3\times M_3}{\mathbb{Z}_2}\,.
\end{equation}
This implies that both 3-manifolds are identical with identical volumes that are simply set by the 6-dimensional volume. The orientifold cycle is given by the formal sum of the two 3-cycles $M_3 + M_3$, and its volume  is directly determined by the overall 6-dimensional volume.
Such examples are non-generic, but the results of this paper will suggest that they are perhaps more useful since fluctuations in the subspace $\rho, \tau,\sigma$ will tend to be unstable. A necessary, but not sufficient, condition for $\sigma$ to be rigid  is to have a $2n$-dimensional internal space with an orientifold that wraps an $n$-dimensional submanifold. If then the orientifold wraps an even combination of $n$-cycles, for which the separate components are mapped to their Hodge dual, under the involution, the orientifold cycle is non-dynamical. Later in this paper we consider $SU(2)\times SU(2)$ with four $O6$ planes, and the above effect will take place in such a way that only three $O6$ planes have dynamical volumes.

\subsection{The scaling of the internal curvature}

The Ricci scalar is schematically of the form
\begin{equation}
R \sim  g^{-1} \partial (g^{-1}\partial g) + g^{-1} (g^{-1}\partial g)^2\,,
\end{equation}
where we suppressed the indices on the metric and its inverse, since many possible contractions arise. From the above expression one expects many scalings to be  possible. Similar to the form fluxes, specific scalings will be forbidden by the orientifold involution symmetry. Let us first discuss this for group manifolds, which are popular in model building since they allow explicit computations, even from a 10D point of view. The covering space of an $n$-dimensional group manifold $G$ is defined by $n$ globally-existing one-forms $\eta^A$, that obey
\begin{equation}
\d\eta^A = -\tfrac{1}{2}f^A_{BC}\eta^B\wedge\eta^C\,,
\end{equation}
where $f^A_{BC}$ are the structure constants of the Lie algebra $\mathfrak{G}$ associated to the group $G$. For such manifolds the orientifold involution is typically defined on these one-forms. One can always choose a basis in which the one-forms $\eta^A$ are split into even $\eta^a, \eta^b, \ldots$  and odd $\eta^i, \eta^j, \ldots$  forms. In other words: the orientifold plane extends along the directions $\eta^a$. The involution symmetry then allows only the following non-zero structure constants
\begin{equation}\label{components}
f^{a}_{bc}\,,\qquad f^{a}_{ij}\,,\qquad f^{i}_{aj}\,.
\end{equation}
The metric on the manifold is given by
\begin{equation}
\d s^2 = g_{AB} \eta^A\otimes \eta^B\,,
\end{equation}
where $g_{AB}$ is positive definite and symmetric and hence contained in $\SL(n, \mathbb{R})/\SO(n)$. The orientifold involution guarantees that the off-diagonal components of the kind $g_{ai}$ are zero. The tangent space metric $g_{AB}$ can only depend on the lower-dimensional coordinates and it effectively contains the left-invariant metric scalar fields. Using that $g_{ab}$ scales as $\sigma^A$ and $g_{ij}$ as $\sigma^B$, we find from the curvature formula
\begin{equation}
R = \tfrac{1}{2}g^{EA}f^B_{CE} f^C_{AB}  + \tfrac{1}{4}g^{LA} g^{BE}
g_{DC} f^D_{EL} f^C_{AB}\,.
\end{equation}
that the only allowed scalings are of the form
\begin{equation}\label{scaling2}
\sigma^{-A}\,,\qquad \sigma^{-B}\,,\qquad \sigma^{-2B + A}\,.
\end{equation}
where we used that the only non-zero components are given by (\ref{components}). This is a significant simplification, since without the involution symmetry, one would expect many more possible scalings.

\subsection{Stability constraints from $\sigma$ fluctuations?}\label{sec:stability}
From the stability analysis in terms of $\rho$ and $\tau$ \cite{Shiu:2011zt, VanRiet:2011yc} it is clear that moduli that scale in a universal way lead to universal stability conditions and are therefore very strong. As explained, $\sigma$ is semi-universal in the sense that the $\sigma$ scalings in the scalar potential are quite universal, except for the Ricci scalar contribution. From this fact alone we expect (and show this below) that many random solutions are unstable already in the 3-dimensional moduli space spanned by $\rho, \tau$ and $\sigma$. However, as we review now there are more reasons to believe that these fluctuations are amongst the most dangerous. Especially when the fluctuations go into the $\sigma$ directions.

There exist models in which the existence of instabilities can be rigourously proven. These models are defined by $N$ real scalars $\Phi_i$  that reside in a scalar potential that has $N+1$ terms which are all of the form
\be \label{exponential}
V =\sum_{a=1}^{N+1} \Lambda_a \exp(\vec{\alpha}_a \cdot \vec{\Phi})\,,
\ee
where each exponential term contains a linear combination of scalars $\vec{\alpha}_a \cdot \vec{\Phi} = \sum_{i=1}^N \alpha_{ai}\Phi^i$.  It has been shown in \cite{Hartong:2006rt} (see Appendix B), that any critical point of such a function with a \emph{positive} value of $V$, has a Hessian with at least one negative eigenvalue, if one of the coefficients $\Lambda_a$ is negative. The unstable modes are then residing in the fields contained in the exponential of the negative term. Interestingly this does not need to occur at negative values of the scalar potential.

Apart from the choice of numbers; $N$ scalars and $N+1$ terms, this potential is of  a completely generic form for tree-level flux compactifications, since we did not specify a certain form for the kinetic term and all possible field redefinitions, with non-zero Jacobian at the critical point,  are allowed to get the potential in this form. For example, in terms of canonically normalised scalar fields one often encounters `axion-dilaton' $\varphi, \chi$ pairs in $\mathcal{N}=1$ flux compactifications, whose kinetic terms look like
\be
\mathcal{L}_{\text{kin}}\sim -(\partial\varphi)^2 -\e^{c\varphi}(\partial\chi)^2
\ee
At tree-level one then finds that $\chi$ appears polynomial in the potential and $\phi$ appears polynomial in its exponential. If we redefine the axion field $\chi = \pm \e^{\tilde{\chi}}$, then both scalars $\phi,\tilde{\chi}$ will only appear exponentially in the scalar potential.  It would be interesting to see what happens in the case when there are more than $N+1$ terms, as is usually the case for flux compactifications. It is to be expected that there cannot be a generic proof for unstable modes anymore but that  constraints can be found that get stronger as the number of terms lowers down to $N+1$.

Let us illustrate this explicitly with the most simple example of a single scalar field $x$. Consider a generic potential, which, after suitable field redefinitions, we can  write in a \emph{polynomial} form. One term, at least has to be positive in order to allow for positive extrema. Let us take that to be the first term. We  then consider a redefinition such that the second term is linear and that the first term has coefficient equal to $1$:
\be
V (x) = x^n - a x\,.
\ee
Where $a>0$ is some coefficient. The critical point is at
\be
x = \lp\frac{a}{n}\rp^{\frac{1}{n-1}}\,,\qquad V =
a^{\frac{n}{n-1}}n^{-\frac{1}{n-1}}\lp\frac{1}{n} - 1\rp\,.
\ee
The second derivative of $V$ is
\be
V'' = a(n-1)\lp\frac{a}{n}\rp^{-\frac{1}{n-1}}\,.
\ee
From these expressions one can easily verify that all critical points with positive values are maxima and those with negative values of $V$ are minima.

In the `simple' tree-level orientifold flux compactifications we consider, dS solutions arise likewise from balancing a series of positive terms $V_R, V_i, V_H$ against the negative term $V_{Op}$.\footnote{Strictly speaking there can be multiple terms in $V_{Op}$ and furthermore, there can be negative terms in $V_R$, as long as the total $V_R$ is positive.} At a given critical point in moduli space, one can imagine that similarly to the case above fluctuations around it can lower the energy if these fluctuations predominantly change $V_{Op}$ such that $V_{Op}$ becomes more negative whereas the changes in $V_R, V_i, V_H$ are less severe. The scalars that reside in the negative term $V_{Op}$ are the universal ones $\rho, \tau$ and all the $\sigma_a$ moduli. Hence one can expect that a certain linear combination of those moduli is likely to be unstable. The fluxes and the curvature are generically also sensitive to these fluctuations but in a less direct way. In some cases the fluxes are not sensitive at all to the $\sigma$ fluctuations, as is the case for all RR fluxes in the $O6$ plane models, as shown above. This is of course no proof, and by hand, one can engineer scalar potentials of the type that arise in tree-level flux compactifications that have meta-stable dS vacua. However, we find that the above intuition applies to all the explicit examples of type IIA flux compactifications we study below.

\section{$O6$ compactification on $\SU(2)\times \SU(2)$ }\label{SU2}

In this section we illustrate the above ideas in a very concrete model of classical dS solutions studied in \cite{Caviezel:2008tf, Danielsson:2010bc, Danielsson:2011au}, which comes from massive IIA on $SU(2) \times SU(2)$, containing four intersecting and space-filling $O6$ planes.  Along the way we discover new interesting features of these solutions, such as a supersymmetric limit and a related link with the sgoldstino tachyons studied by many authors before, see for instance \cite{GomezReino:2006dk, Covi:2008ea, GomezReino:2008bi, Borghese:2010ei} and references therein. The model is a particularly nice playground since the dS solutions can be understood as a rather simple 10-dimensional solution \cite{ Danielsson:2010bc}. This means we do not need to rely on numerics to understand the existence of the solution. It furthermore allows one to study the effects of flux quantisation on the set of solutions \cite{Danielsson:2011au}. The simplicity of the solution and its simple 10-dimensional lift  originates from an `isotropy property' of the solution. In simple terms this means that all of the $\sigma_a$ variables can be identified due to an extra $\mathbb{Z}_3$ symmetry, present on-shell. In the appendix we also prove that solutions with this property are unique to $\SU(2)\times\SU(2)$.

\subsection{The geometry and the $\sigma$ moduli}\label{SU2SU2}

We use the following basis to describe the Lie algebra of $SU(2)\times SU(2)$ \cite{Danielsson:2010bc}
\be
f^1{}_{23} = f^1{}_{45} = f^2{}_{56} = -f^3{}_{46} = 1\,, \text{ cyclic.}
\ee
To go to an $\mathcal{N}=1$ supergravity theory in 4 dimensions, we need sufficient orientifolding. The algebra turns out to allow enough $\mathbb{Z}_2$ symmetries for the following BPS intersection of $O6$ planes
\begin{center}
  \begin{tabular}{|c|c|c|c|c|c|}
    \hline
    \rule[1em]{0pt}{0pt} $\eta^1$ & $\eta^2$  & $\eta^3$  & $\eta^4$ & $\eta^5$  & $\eta^6$  \\
    \hline
    \hline
    \rule[1em]{0pt}{0pt} $\bigotimes$ & $\bigotimes$  & $\bigotimes$ & -- & -- & -- \\\hline
    \rule[1em]{0pt}{0pt} $\bigotimes$ & -- & -- & $\bigotimes$  & $\bigotimes$ & -- \\\hline
    \rule[1em]{0pt}{0pt} -- & $\bigotimes$ & -- & -- & $\bigotimes$  & $\bigotimes$ \\\hline
    \rule[1em]{0pt}{0pt} -- & -- & $\bigotimes$ & $\bigotimes$ & -- & $\bigotimes$ \\\hline
   \end{tabular}
\end{center}
The K\"ahler form and the holomorphic 3-form are given in terms of the moduli $a, b, c$ (``K\"ahler'') and $v_1, v_2, v_3, v_4$ (``complex structure'') as follows
\begin{align}
& J = a\eta^{16}+ b\eta^{24} + c\eta^{35}\,,\\
&\Omega_R =v_1\eta^{456} + v_2\eta^{236} + v_3\eta^{134} + v_4\eta^{125}\,.
\end{align}
This defines the following diagonal metric on the tangent space
\begin{equation}
g=\frac{1}{\sqrt{v_1v_2v_3v_4}}\Bigl(a v_3v_4\,,\, -bv_2v_4\,,\,
cv_2v_3\,,\, -bv_1v_3\,,\, cv_1v_4\,,\, av_1v_2\Bigr)\,.
\end{equation}
The standard normalisation requires that \cite{Danielsson:2010bc}
\begin{equation}
\sqrt{v_1v_2v_3v_4}=-abc = \text{Vol} = \sqrt{g} =\rho^3\,.
\end{equation}
From the metric expression it is straightforward to compute the
volumes of the different $O6$ cycles:
\begin{align}
& \text{Vol}(e_1 e_2 e_3) = \sqrt{v_1^{-1}v_2 v_3 v_4} = \rho^{3/2}\sigma_1^{-9/2}(\sigma_2\sigma_3\sigma_4)^{3/2} \,,\label{relation1}\\
& \text{Vol}(e_1 e_4 e_5) =\sqrt{v_1 v_2^{-1} v_3 v_4 }=    \rho^{3/2}\sigma_2^{-9/2}(\sigma_1\sigma_3\sigma_4)^{3/2} \,,\\
& \text{Vol}(e_2 e_5 e_6) = \sqrt{v_1 v_2 v^{-1}_3 v_4} =  \rho^{3/2}\sigma_3^{-9/2}(\sigma_1\sigma_2\sigma_4)^{3/2} \,,\\
& \text{Vol}(e_3 e_4 e_6) = \sqrt{v_1 v_2 v_3v^{-1}_4} = \rho^{3/2}\sigma_4^{-9/2}
(\sigma_1\sigma_2\sigma_3)^{3/2} \,,\label{relation4}
\end{align}
where we defined the moduli $\sigma_a$, $a =1\ldots 4$, as in equation (\ref{sigma}). Relations (\ref{relation1}-\ref{relation4}) should be regarded as the relations that define $\rho, \sigma_a$ in terms of the $v_i$. Clearly this is an under determined algebraic system and we can find infinitely many solutions. So we can take a ``gauge'' to our liking, for instance we can set one of the $\sigma_a$ to unity.  Hence, there are only three-independent dynamical volumes. This is consistent with what we described around equation (\ref{dynamical}).

\subsection{The isotropic solution}\label{sec:10Disotropic}
The de Sitter solutions that can be found in this geometry contain a simple class that can be treated analytically from a ten-dimensional point of view. These solutions are a bit simpler than the typical de Sitter solutions because they are more ``isotropic'' \cite{Danielsson:2010bc}. These solutions are based on half flat $\SU(3)$ structures, with $W_2=0$. The Ansatz for the non-zero fluxes is then
\begin{align}
e^{\Phi}F_0  & = f_1 \, , & e^{\Phi}F_2 &= f_2 J \, , \hspace{3cm} \label{eq:TORSION1}\\
H & = f_5 \Omega_R + f_6 \hat{W}_3 \, , & e^{\Phi}j  &= j_1 \Omega_R
+ j_2 \hat{W}_3 \, ,\label{eq:TORSION2}
\end{align}
where we define the normalized torsion class
\begin{equation}
\hat{W}_3 = \frac{W_3}{\sqrt{|W_3|^2}}\,,
\end{equation}
and $j$ is the source 3-forms for the $O6$ planes. The above is a very simple Ansatz in which $F_4$ and $F_6$ vanish (which correspond to $f_3=f_4=0$ in the language of \cite{Danielsson:2010bc}).

The geometric moduli values for the isotropic $\SU(2)\times\SU(2)$ compactifications are given by
\begin{equation}\label{isotropy}
a=-b=c\,,\qquad v_1=v_2=v_4\equiv v\,,\qquad v_3=\frac{a^6}{(v)^3}\, .
\end{equation}
This condition is the reason for the name ``isotropic''  and one can show it is consistent with having $W_2=0$. For these values we have explicitly
\begin{align}
g & = \text{diag}\left(\tfrac{a^4}{(v_1)^2},\tfrac{(v_1)^2}{a^2},\tfrac{a^4}{(v_1)^2},\tfrac{a^4}{(v_1)^2},\tfrac{(v_1)^2}{a^2},\tfrac{(v_1)^2}{a^2}\right) \, , \\
W_1 & = \frac{a^6 + (v_1)^4}{4 \, a^5 v_1}  \, , \\
W_3 & = \frac{a^6-3 (v_1)^4}{8 \,a^5 (v_1)^4} \left[ (v_1)^4
(e^{456}+e^{236}+e^{125})-3 \,a^6 e^{134}\right]  \, ,
\end{align}
For certain regions in parameter space $f_1, f_2, j_1, j_2$ there exist explicit de Sitter solutions for specific values of $\rho$, $\tau$ and $\sigma$ \cite{Danielsson:2010bc}.\footnote{Flux and charge quantisation removes most, if not all, of the solutions \cite{Danielsson:2011au}.}

For the isotropic case we have $v_1 = v_2 = v_4$ which implies $\sigma_1 = \sigma_2 = \sigma_4$ so that we have only two $\sigma_a$. As explained below \eqref{relation1}-\eqref{relation4} these two $\sigma_a$ are not independent and we can choose a handy gauge.\footnote{The scaling of all the terms in the potential is in this case gauge dependent. We choose our gauge such that $V_H$ scales as in \eqref{eq:Hscaling}.} If we set
\begin{equation}\label{eq:O6scaling}
\sigma_1 = \sigma_2 = \sigma_4=1\,, \quad\sigma_3 \equiv \sigma\,,
\end{equation}
we find
\begin{equation}
a =\rho\,,\qquad \sigma=\rho\, v^{-2/3}\,.
\end{equation}
At the isotropic point the metric is given by
\begin{equation}
g=\rho\,\text{diag}(\sigma^{3}, \sigma^{-3}, \sigma^{3}, \sigma^{3}, \sigma^{-3}, \sigma^{-3})\,.
\end{equation}
One can verify \cite{Danielsson:2010bc} that there is an interval of de Sitter solutions for the values:
\begin{equation}
4.553 <\frac{w_3}{W_1}<3\sqrt3\,.
\end{equation}
The two boundaries of the interval correspond to Minkowski solutions. For $4.553 >w_3/W_1$ we find AdS solutions instead. The other end of the interval $w_3/ W_1=3\sqrt3$ cannot be exceeded as it corresponds to a boundary of moduli space for which
\begin{equation}
v^4/a^6\rightarrow \infty .\label{limit}
\end{equation}
What has not been remarked in \cite{Danielsson:2010bc} is that this limit can be taken such that it leads to a supersymmetric but highly degenerate solution. This is very interesting because, as we show below, near the supersymmetric Minkowski solution the tachyonic direction is the sgoldstino direction and we are able to analytically identify the tachyon for a certain range of parameters. The conditions for an $\mathcal{N}=1$ Minkowski vacuum in IIA supergravity with $O6$ planes can be expressed in terms of first-order equations for the pure spinor polyforms \cite{Grana:2004bg, Grana:2005sn} and boils down to three conditions
\begin{align}
& \d \Omega_R =0 \,,\label{SUSY1}\\
& \d J =0\,, \label{SUSY2}\\
& \d \Omega_I = g_s\star F_2 \,, \label{SUSY3}
\end{align}
and all fluxes, but $F_2$ are zero:
\begin{equation}
H, F_0, F_4 , F_6\rightarrow 0\,.
\end{equation}
This also implies that the $O6$ planes wrap cycles that are trivial in homology.
Since $\d J = \tfrac{1}{2} a (e^{456}+e^{236}+e^{125})$, supersymmetry requires that the limit (\ref{limit}) has to be taken such that the volume goes to zero
\begin{equation}
a\rightarrow 0 \,,
\end{equation}
so that this limit does not lead to a trustworthy supergravity solution. However, as we will argue below it allows us to analytically identify the tachyonic direction near this SUSY Minkowski point.

Due to the choice of a half-flat $\SU(3)$-structure on $\SU(2)\times \SU(2)$ we automatically obey $\d\Omega_R=0$. Therefore only (\ref{SUSY3}) remains to be checked. We furthermore have that, at the Minkowski limit, $f_2 = 2W_1$ \cite{Danielsson:2010bc}. Equation (\ref{SUSY3}) then simply follows from the $\SU(3)$-structure identities (with $W_2=0$) and \eqref{eq:TORSION1}
\begin{equation}
\d \Omega_I = W_1 J \wedge J\,,\qquad \star_6 J=\tfrac{1}{2}J\wedge J\,.
\end{equation}

\subsection{The scalar potential for the isotropic solution}\label{sec:Visotropic}

As we have shown above in subsection \ref{sec:10Disotropic}, the one-parameter family of solutions for the isotropic case has a limiting point that can be understood as a (highly degenerate) supersymmetric Minkowski vacuum. Since supersymmetric solutions have generically a much simpler structure we study this limiting point in detail in the next subsection and find the explicit direction of the tachyon near this limiting point.

We recall that the scalar potential arises as F-term potential in an $\mathcal{N}=1$ supergravity theory and is therefore given by (see \cite{Danielsson:2011au} and references therein)
\be\label{VFterm}
V=e^K (K^{ij} D_i W \overline{D_j W} -3 |W|^2),
\ee
with
\ba
K &=& -\log(z_1+\bar{z}_1)-3\log(z_2+\bar{z}_2) -3\log(t+\bar{t}) + 5 \log(2)\,, \\
W &=& i \lambda t^3+3 t (t+z_1+z_2)-i \lambda (z_1-3 z_2)\,,
\ea
where we have introduced the complex moduli
\be
t=\rho - i b\,,\quad z_1 = \tau \sigma^{-9/2} + i c_1\,, \quad z_2 = \tau \sigma^{3/2} + i c_2,
\ee
with $b$ being the axion arising from $B_2$ and the $c_i$ are axions arising from $C_3$. These three complex scalars corresponds to the so-called STU truncation \cite{Dibitetto:2011gm}.

We have rescaled the moduli fields and the scalar potential to set the $F_0$ and $F_2$ flux parameters to unity. The parameter $-8.7 \lesssim \lambda \lesssim -4.8$ that appears in $W$ corresponds to the $H$ flux and is the one free parameter for our dS solutions. $\lambda \approx -8.7$ corresponds to the highly degenerate, supersymmetric Minkowski vacuum. What this means in terms of the four dimensional potential is that one could rescale the moduli and fluxes such that $W$ and all the $F_i = D_i W$ vanish. However, since this highly degenerate Minkowski limit has vanishing volume one finds that $e^K \sim 1/$Vol diverges. We are only using the existence of this special point as a hint at which region in moduli space might be easy to understand. Therefore we have chosen a different rescaling of the moduli to fix the RR flux parameters to unity since this leads to nice F-term behavior that we discuss in the next subsection. For our choice of rescalings one finds that near $\lambda \approx -8.7$, $e^K$ diverges while $K^{ij} D_i W \overline{D_j W} -3 |W|^2$ vanishes leading to $V \rightarrow \infty$ near this `degenerate, supersymmetric Minkowski' point.

The explicit contributions to the scalar potential \eqref{VFterm} are
\be
V = V_H + V_R + V_{O6} + V_0 +V_2\,,
\ee
which become very simple after we have solved $\partial_{c_1} V = \partial_{c_2} V=0$ which leads to
\be
c_1=\frac{b (3 (b-2) \lambda +b (2 b-3))}{4 \lambda }\,,\quad c_2=\frac{b (b (\lambda +3-2 b)-2 \lambda )}{4 \lambda }\,.
\ee
The contributions are then
\ba\label{eq:scalarpotential}
V_H &=& \frac{1}{\rho^{3} \tau^{2}} \lp\frac{(\lambda-3b)^2}{\sigma^{9}}+3 (\lambda+b)^2 \sigma^3 \rp\,,\quad  V_R = \frac{3}{\rho\,\tau^{2}} \lp \frac{1}{\sigma^{9}}-\frac{4}{\sigma^3}-\sigma^3\rp\,,\\
V_{O6} &=& \frac{2}{\tau^{3}} \left(\frac{\lambda -3}{\sigma^{9/2}} -3 (\lambda +1) \sigma ^{3/2}\right)\,, \quad V_0=\frac{\rho^3}{\tau^4}\,,\quad V_2 = \frac{3 \rho (b -1)^2}{\tau^4}\,,
\ea
and we see that $V_H$, $V_R$ and $V_i$ exhibit the expected scalings with $\rho$, $\tau$ and $\sigma$ as derived in \eqref{eq:Hscaling}, \eqref{eq:RRscaling}, \eqref{eq:AB}, \eqref{scaling2}. The scaling for $V_{O6}$ follows from \eqref{relation1}-\eqref{relation4} and the handy gauge choice \eqref{eq:O6scaling}.

\subsection{The sgoldstino direction}
In this subsection we are studying minors of the mass matrix of the isotropic $SU(2) \times SU(2)$ model and check whether they have a negative eigenvalue. By identifying simple minors one can hope to ultimately get an analytic handle on the tachyon and maybe even generalize the result to other models.\footnote{A negative eigenvalue in a minor implies a negative eigenvalue for the entire mass matrix due to Silvester's criterion.}

The first obvious choices for directions along which to (numerically) calculate the minor of the mass matrix are the two universal moduli $\rho$ and $\tau$. However, it turns out that the $2\times2$ minor spanned by these two directions does never contain the tachyon so that we need to study further directions. Using our new additional universal modulus $\sigma$ we find that the $2\times2$ minor spanned by $\rho$ and $\sigma$ never contains a tachyon but the $\tau$, $\sigma$ minor does contain the tachyon for $-8.7 \lesssim \lambda \lesssim -5.4$. Finally, the $3 \times 3$ minor spanned by all our three (semi-)universal moduli $\rho$, $\tau$ and $\sigma$ always contains the tachyon for the entire range of the one-parameter, demonstrating the usefulness of our additional semi-universal modulus $\sigma$.

Unfortunately we are not able to obtain the analytic solution for the one-parameter family of dS critical points since they are determined by the root of an irreducible polynomial of degree 19 \cite{Danielsson:2011au}. For that reason we cannot get an analytic expression for the tachyonic direction. However, having established the existence of the degenerate, supersymmetric limiting Minkowski point, we can study the dS solutions near this point. While we are not claiming that these solutions are trustworthy, we find that the tachyon takes a very simple form. Near that point the F-terms $F_t$ and $F_{z_1}$ are going to zero while the F-term $F_{z_2}$ diverges. This means we should focus on $z_2$ and indeed we find that the tachyon is given by the sgoldstino direction and is along the $\text{Re}(z_2) = \tau \sigma^{3/2}$ direction. In particular for $-8.7 \lesssim \lambda \lesssim -6.1$ we find at the extremum that $\partial_{\text{Re}(z_2)} \partial_{\text{Re}(z_2)} V<0$. The first term of $V_{O6}$ in \eqref{eq:scalarpotential} couples only to $\text{Re}(z_2)^{-3} = \tau^{-3} \sigma^{-9/2}$ so that there is a clear relation between the tachyonic direction and this negative term in the scalar potential. Near this limiting point the first term of $V_{O6}$ in \eqref{eq:scalarpotential} is the dominant negative contribution so it is tempting to think that some generalization of the no-go theorem reviewed in subsection \ref{sec:stability} is at work.

\section{More examples of massive IIA supergravity}\label{scan}

Having shown the usefulness of the $\sigma$ direction in the explicit case of the isotropic $SU(2)\times SU(2)$ example, one might wonder whether the same holds for other examples. In this section we study many more dS critical points and find further evidence that relate the $\rho$, $\tau$ and $\sigma_a$ directions to the tachyonic direction.

In \cite{Danielsson:2011au} a systematic search for dS vacua in massive type IIA flux compactifications was performed.\footnote{A different scan for tree-level compactifications is given by $\mathcal{N}$-extended gauged supergravities, for which there exists an extended body of literature that discusses de Sitter solutions. Meta-stable dS solutions have only been found for $\mathcal{N}\leq 2$, see for instance \cite{Fre:2002pd, Roest:2009tt}, but the higher-dimensional origin of these models still has to be found. Very recently progress has been made for the case $\mathcal{N}=8$ \cite{Dall'Agata:2012sx }, but also here the higher dimensional origin is not yet clear.} The authors studied all SU(3)-structure manifolds that can be realized as orbifolds of a group or coset manifold. They found more than a dozen of explicit group manifolds that can give rise to dS critical points. Since all of these critical points had at least one tachyonic direction, they are a natural playground to check the relevance of the $\sigma_a$ directions.

All of these compactifications have 2 or 3 K\"ahler moduli with 2 or 3 corresponding $B_2$ axions as well as the dilaton and three complex structure moduli with the corresponding four $C_3$ axions. The three complex structure moduli are again related to three independent $\sigma_a$ similarly to the $\SU(2)\times \SU(2)$ case discussed in subsection \ref{SU2SU2}. We thus have more than a dozen models with twelve or fourteen real moduli. These moduli contain the universal moduli $\rho$ and $\tau$ corresponding to the overall volume and the dilaton as well as three semi-universal $\sigma_a$. Since it is time consuming to numerically find dS critical points, we have not been able to map out the entire space of critical points for any of the more complicated models. However, for the more than a dozen examples we have calculated in total over one hundred critical points. We found that the tachyonic direction is often contained in the universal $2\times2$ minor spanned by $\rho$ and $\tau$ however in roughly 10\% of the dS critical points this was not the case. We found however that the tachyon is always contained in the $5\times5$ minor spanned by $\rho$, $\tau$ and the three $\sigma_a$. This strongly indicates that the $\sigma_a$ are very important for stability as well. One might have expected that the axionic directions might be less relevant for stability but in our examples we are also finding that out of all the K\"ahler moduli only the overall volume seems to be relevant for stability. It would be interesting to make this observation more precise.

\section{Discussion}\label{discussion}
In this note we have tried to answer why all known tree-level orientifold compactifications with fluxes give rise to perturbatively unstable de Sitter solutions. There is no conclusive answer and it therefore remains an excellent challenge to try to construct perturbative stable solutions. However the results of this paper give some clues on where to look for meta-stable solutions. As we have shown, strong constraints for meta-stability arise from those field directions that scale universal (the dilaton and overall volume) and semi-universal (the orientifold volumes) in the scalar potential. We have verified for all known cases that the instability is in those directions. Therefore prior to any detailed computation of the mass spectrum for a given compactification it might be more efficient to first check the fluctuations around a solution in that subspace of field space.

A further clue is given by the arguments of random potentials \cite{Chen:2011ac, Marsh:2011aa, Sumitomo:2012wa, Bachlechner:2012at, Sumitomo:2012vx, Sumitomo:2012cf } that clearly all have in common that models with less moduli have a better chance of giving meta-stable solutions. From the latter point of view it might be beneficial to study flux compactifications to 5 or 6 dimensions, since these compactifications can still evade the simplest no-go theorems against meta-stable dS and have fewer moduli \cite{VanRiet:2011yc}.

In the case that SUSY is broken by a small amount, compared to the usual masses of the scalar fields in the supersymmetric vacuum, there also exist powerful and useful constraints for the meta-stability of dS solutions, based on the sgoldstino directions \cite{GomezReino:2006dk, Covi:2008ea, GomezReino:2008bi, Borghese:2010ei}. For tree-level flux compactifications one generically expects no fine tunings to be possible and SUSY is generically broken close to the KK scale. Nonetheless we have shown that for the $\SU(2)\times\SU(2)$ compactification of massive IIA there exist an (albeit singular) SUSY limit in which the tachyon indeed coincides with the sgoldstino direction. It furthermore turns out that the sgoldstino near that point coincides with the orientifold volume, which relates two approaches for understanding the structural reasons for instabilities in the context of tree-level type IIA flux compactifications.

\section*{Acknowledgments}
We like to thank  Sujan Dabholkar, Giuseppe Dibitetto, Daniel Junghans, Savdeep Sethi and Paul Smyth for useful discussions. U.H.D. is supported by the G{\"o}ran Gustafsson Foundation and the Swedish Research Council (VR). During most stages of this research TVR was supported by the ERC Starting Independent Researcher 259133-ObservableString, and currently TVR is supported by a Pegasus Marie Curie fellowship of the FWO. The work of GS was supported in part by DOE grant DE-FG-02-95ER40896. GS would also like to thank the University of Amsterdam for hospitality during part of this work, as he was visiting the Institute for Theoretical Physics as the Johannes Diderik van der Waals Chair. The work of T.W. was supported by a Research Fellowship (Grant number WR 166/1-1) of the German Research Foundation (DFG).

\appendix

\section{Isotropic dS on more general manifolds? }
The 10-dimensional lift of the four-dimensional isotropic dS solutions of section \ref{sec:10Disotropic} are written entirely in terms of the non-zero torsion classes $W_1, W_3$ and the $\SU(3)$-structure forms $J$, $\Omega$ \eqref{eq:TORSION1}, \eqref{eq:TORSION2}. This approach was followed in \cite{Danielsson:2009ff, Danielsson:2010bc, Andriot:2010ju} and it mimics the form of supersymmetry-preserving vacuum solutions \cite{Lust:2004ig, Grana:2004bg, Grana:2005sn, Grana:2006kf}. This approach has the advantage that it does not require a specific choice of manifold, instead it only requires relations among the torsion classes and the $\SU(3)$-invariant forms $J$, $\Omega$. These relations could then be satisfied by whole families of manifolds. The fact that solutions can be established without choosing a specific manifold relies on the fact that the 10D Ricci tensor, as required for the Einstein equations, can be written entirely in terms of the torsion forms and $J$ and $\Omega$ \cite{bedulli-2007-4}. If one goes through the 10D Einstein equations and the form field equations, one finds a set of restrictions on the $\SU(3)$-structure, which can be interpreted as fixing the moduli to specific values. In order to present these restrictions we first need to define two tensors $Q_1, Q_2$ as in \cite{Danielsson:2010bc}
\begin{align}
& Q_1 = \Bigl(\Omega^{ijk} \iota_j \iota_i \hat{W}_3 \wedge \iota_k \hat{W}_3 \Bigr)|_{(2,1)}\,,\label{Q1}\\
& Q_2 =\Bigl( \tfrac{1}{2}\hat{W}_{3imn}\hat{W}_3^{pmn}\Omega_{pjk} e^i\wedge e^j\wedge e^k\Bigr)|_{(2,1)}\,. \label{Q2}
\end{align}
Our notation is such that the subscript on the expression means we project to the $(2,1)$-part. Secondly the hatted $\hat{W}_3$, denotes the normalised $W_3$:
\be
W_3=\sqrt{ (w_3)^2}\hat{W_3}\,,\qquad w_3^2 =\tfrac{1}{3!}W_{3abc}W_3^{abc}\,.
\ee
The restrictions on the torsion classes, enforced from the equations of motion, are then
\begin{align}
& \d\star_6 \hat{W}_3 = c_1 J\wedge J\,,\label{res0} \\
& (\hat{W}_{3i}\cdot\hat{W}_{3j})^+ = 0\,,\label{res1}\\
& Q_1 = c_2 Q_2=c_3 (W_3)_{2,1}\,,\label{res2}
\end{align}
where $c_1, c_2, c_3$ are some real numbers and $ (\hat{W}_{3i}\cdot\hat{W}_{3j})^+$ is that part of the symmetric tensor $ (\hat{W}_{3i}\cdot\hat{W}_{3j})$ that transforms in the $\bold{8}$ of $\SU(3)$. From the definitions of the torsion classes one can deduce that we necessarily have to fix $c_1$ to $c_1 =w_3/6$.  We will prove this below.  In \cite{Danielsson:2010bc} it was then shown that for $\SU(2)\times \SU(2)$ we must have that
\be
c_2 =1 \,,\qquad c_3=\frac{8}{\sqrt3}\,,
\ee
but it was not understood whether other values would be possible for different manifolds. In what follows we demonstrate that the values for $\SU(2)\times \SU(2)$ must be the same for all manifolds.

Let us first demonstrate that $c_1= w_3/6$. We start with the restricted $\SU(3)$-structure that we have assumed
\be
\d J = \tfrac{3}{2}W_1 \Omega_R + W_3\,,\qquad \d \Omega_R= W_1 J\wedge J\,,
\ee
and then we use the Leibniz rule
\be
(\d \star W_3 )\wedge J= \d (\star W_3 \wedge J) + \star W_3 \wedge \d J\,.
\ee
The RHS of this equation equals $w_3^2\text{Vol}$ by virtue of the $\d J$ identity and the fact that $\star_3 W\wedge J=0$ (which is a consequence of the complexity types of $J$ and $W_3$.) The LHS of the equation is $6 c_1 w_3 \text{Vol}$ since we have chosen the normalisation $J \wedge J\wedge J= 6\text{Vol}$.

Now we demonstrate that $c_2=1$. This was known for $\SU(2)\times\SU(2)$, but it turns out to be true generally. Once we assume the restrictions \eqref{res1}, \eqref{res2} then we can contract both $Q_1$ and $Q_2$ with $W_3$ and we can drop the $(2,1)$ subscript since the projection on that part is then automatic. This gives
\begin{align}
& (Q_1)_{abc}W^{abc} \sim \Omega^{ijk}W_{jia}W_{kbc}W^{abc}\,,\nonumber \\
& (Q_2)_{abc}W^{abc} \sim  \Omega^{pbc}W_{pmn}W_{abc}W^{amn}\,,
\end{align}
which after relabeling is exactly the same expression. Hence $c_2$ must always be the same expression. If $c_2=1$ for $\SU(2)\times \SU(2)$ then it is always equal to $1$. It seems very reasonable that the same type of computation allows one to prove that $c_3$ is always constant, but instead we will get this result by interpreting the 10D equations of motion as the equations for a 4D vacuum solution
\be
\partial_i V(\phi)=0\,, \qquad R_4 = 2V(\phi)\,,
\ee
where $V$ is the scalar potential and $\partial_i$ is the derivative with respect to the various scalars one gets from the dimensional reduction. As shown in \cite{Danielsson:2010bc} our 10D Ansatz in terms of the forms $W_1, W_3, J, \Omega$ gives seven algebraic equations, once the restrictions \eqref{res0}, \eqref{res1}, \eqref{res2} are assumed. This would imply that there exists a truncation down to six real moduli of the generic model with the specific $\SU(3)$- structure constraints described above. This truncation was noted in \cite{deCarlos:2009qm} for the $\SU(2)\times \SU(2)$ model and it originates from an extra $\mathbb{Z}_3$ symmetry that arises when some of the fluxes are set to zero. This extra symmetry can be understood as the isotropy condition (\ref{isotropy}). On the level of the 4D $\mathcal{N}=1$ supergravity this means that we can consider a minimal model in terms of three complex scalars $S, T, U$ \cite{deCarlos:2009qm, Danielsson:2011au}. The real parts of these three scalars are related to the universal moduli $\rho, \tau$ and $\sigma$. The imaginary parts, called axions,  originate from the gauge potentials. The seven 10D equations of \cite{Danielsson:2010bc} can thus exactly be identified with the six stabilisation equations $\partial_i V=0$ and the definition of the 4D cosmological constant $ R_4 = 2V(\phi)$. Among these seven equations, one can readily recognize the stabilisation with respect to $\tau$ and $\rho$, since these are certain linear combinations of the traced internal Einstein equation and the dilaton equation. Then there are three form field equations for the fluxes, which should be interpreted as the stabilisation equations for the three axions. What remains to be interpreted is an off-diagonal  internal Einstein equation, labeled equation (22g) in \cite{ Danielsson:2010bc}
\begin{equation}\label{22g}
2W_1w_3 -2j_2 -4f_5f_6 -\tfrac{1}{4}w_3^2 c_3 -\tfrac{1}{4}\frac{c_3}{c_2}f_6^2 =0\,.
\end{equation}
One can verify for $\SU(2)\times\SU(2)$ that this equation exactly corresponds to $\tfrac{2}{3\sqrt3}\sigma\partial_{\sigma}V=0$. However, since the $j_2$-term (descending from $V_{O6}$) and the $-4f_5f_6$-term (descending from $V_H$) in (\ref{22g}) are independent of the values of $c_3$ we always have that equation (\ref{22g}) corresponds to $\tfrac{2}{3\sqrt3}\sigma\partial_{\sigma}V=0$. This fixes then $c_3 = \frac{8}{\sqrt3}$, otherwise the $f_6^2$-term cannot  descend from $\partial_{\sigma}V_H$. This implies for instance  that
\begin{equation}\label{condition}
-\frac{2}{3\sqrt3}\sigma \partial_{\sigma}V_R = 2 W_1 w_3 -\frac{c_3}{4}w_3^2\,,
\end{equation}
where  $2V_R = -15 W_1^2 + w_3^2$.

\bibliography{refs}

\bibliographystyle{utphysmodb}

\end{document}